# On the Testing of Ground-Motion Prediction Equations against Small-Magnitude Data

C. Beauval, H. Tasan, A. Laurendeau, E. Delavaud, F. Cotton, Ph. Guéguen, and N. Kuehn

corresponding author:

Céline Beauval

ISTerre

Bp 53

38041 Grenoble Cedex 9

FRANCE

Celine.Beauval@obs.ujf-grenoble.fr

Tel: 00 33 (0)4 76 63 52 08

**Abstract**

Ground-motion prediction equations (GMPE) are essential in probabilistic seismic hazard studies for estimating the ground motions generated by the seismic sources. In low seismicity regions, only weak motions are available in the lifetime of accelerometric networks, and the equations selected for the probabilistic studies are usually models established from foreign data. Although most ground-motion prediction equations have been developed for magnitudes 5 and above, the minimum magnitude often used in probabilistic studies in low seismicity regions is smaller. Desaggregations have shown that, at return periods of engineering interest, magnitudes lower than 5 can be contributing to the hazard. This paper presents the testing of several GMPEs selected in current international and national probabilistic projects against weak motions recorded in France (191 recordings with source-site distances up to 300km, $3.8 \leq M_w \leq 4.5$). The method is based on the loglikelihood value proposed by Scherbaum et al. (2009). The best fitting models (approximately $2.5 \leq LLH \leq 3.5$) over the whole frequency range are the Cauzzi and Faccioli (2008), Akkar and Bommer (2010) and Abrahamson and Silva (2008) models. No significant regional variation of ground motions is highlighted, and the magnitude scaling could be predominant in the control of ground-motion amplitudes. Furthermore, we take advantage of a rich Japanese dataset to run tests on randomly selected low-magnitude subsets, and check that a dataset of ~190 observations, same size as the French dataset, is large enough to obtain stable LLH estimates. Additionally we perform the tests against larger magnitudes (5-7) from the Japanese dataset. The ranking of models is partially modified, indicating a magnitude scaling effect for some of the models, and showing that extrapolating testing results obtained from low magnitude ranges to higher magnitude ranges is not straightforward.

**Introduction**

A probabilistic seismic hazard analysis (PSHA) is usually carried out to establish a national seismic building code. Such analysis relies on the identification of the seismic sources (size, location and occurrence probability) and the estimation of their capacity to produce ground-motions. A ground-motion prediction equation (GMPE) is necessary for estimating the conditional probability that, if the earthquake occurs, a given acceleration threshold can be exceeded at the site of interest. The minimum magnitude used in PSHA studies varies from 5 down to 4, and even lower values in very low seismicity regions such as Scandinavian countries. Desaggregation studies show that the whole range of magnitudes considered contributes to the PSHA, with the barycentre of the contributions depending on the return period considered and the seismicity level of the region. Beauval et al. (2006, 2008) showed that in the active parts of France, magnitudes contributing the most at 475 years are in the interval [5-5.5], but magnitudes 4 to 5 are also responsible for a non-negligible contribution to the hazard even for return periods as large as 10,000 years. Therefore, there is a need for reliable predictions of ground-motion amplitudes over the whole magnitude range.

To develop a robust GMPE, a large accelerometric dataset is required containing a wide range of magnitudes and source-to-site distances. In low-seismicity regions such as France, the bulk of the data consists in low magnitude recordings ($M_w<4.5$). Several studies showed that, due to magnitude-scaling problems, equations based on low-magnitude datasets are not able to correctly predict the ground motions of moderate-to-large magnitudes ($M_w \geq 5$, see Youngs et al. 1995, Bommer et al. 2007, Cotton et al. 2008). The solution proposed up to now is to select the GMPEs among published equations established from strong motions recorded in higher seismicity regions (either global or region-specific models, Bommer et al. 2010). The large majority of these equations have been developed for magnitude five and above (Douglas, 2011). This is the case for all GMPEs selected within the SHARE project for application in crustal active regions (see Delavaud, Cotton et al., 2012). The main objective of the SHARE (Seismic Hazard Harmonization in Europe) project is to provide a seismic hazard model for the Euro-Mediterranean region. In PSHA studies led in low-seismicity regions, the use of models based on allogeneous data and developed for magnitude $M_w \geq 5$ rely on two assumptions: 1) for the same magnitude and same distance, the ground motions produced do not vary much from one shallow active region to the other; 2) the models established from moderate-to-large magnitudes have the ability to predict amplitudes of motions produced by lower magnitudes.

The first assumption, the regional variability of ground motions is a currently strongly debated issue, mainly due to the current lack of data. The issue will be completely solved only when local models will be available in all crustal regions; however this will not happen in a near future (Stafford et al., 2007). Some authors believe that ground motions do not vary much regionally, at least for moderate-to-large magnitudes, as long as the same tectonic environment is considered (Bommer 2006; Stafford et al. 2007). In fact, all global models based on a database including data from different regions of the world are assuming that ground motions are not regionally dependent (e.g. the NGA models, Next Generation Attenuation models, Abrahamson et al. 2008). On the opposite, some authors have highlighted significant regional dependence (e.g. Luzi et al. 2006 for moderate magnitudes in Italy), however often based on restricted regional datasets. For low magnitudes, different recent publications show statistically significant regional differences in ground motions (e.g. Atkinson and Morrisson 2009, Chiou et al. 2010, or Bakun and Scotti 2006 based on macroseismic intensities), while others did not evidence such discrepancy (Douglas 2004). Atkinson and Morrison (2009) and Chiou et al. (2010) found that ground-motion amplitudes from small earthquakes in northern California are lower on average than those for southern California. They observed that these differences are no longer significant for magnitudes larger than ~6. The ground motions from smaller earthquakes may be more sensitive to differences in crustal structure, or in crustal stress states. As stress drops of small earthquakes show to be magnitude-dependent, the regional dependence of the average stress drop could result in different ground motions (Chiou et al. 2010). Besides, as stress is dependent on depth, the focal depth of small magnitude earthquakes might also play a significant role in the scaling of ground motions.

The second assumption resides in the way the models account for the scaling of ground motions with magnitude. Following Bommer et al. (2010) selection criteria, a model should include non-linear scaling of ground-motion amplitudes with magnitude. If the $w^{-2}$ source model is assumed, the corner period of the spectrum varies with the magnitude, and the scaling law of the source spectrum amplitude becomes a non-linear function of magnitude (Fukushima, 1996). The scaling relation of the spectrum amplitude with $M_w$ can be approximated as a quadratic function, and the coefficient of the $M^2$ term should be negative, implying that the rate of decrease in spectral amplitude with decreasing magnitude is accentuated (e.g. Zhao et al. 2006, Bindi et al. 2009). However, these constraints on the magnitude scaling do not enable the extrapolation of GMPEs at the limits or beyond their range of applicability (Bommer et al. 2007). This inability has different potential origin: the magnitude scaling of ground motions that decreases with increasing magnitude, the stronger decay of small-magnitude motions with respect to larger-magnitude motions, and/or the scaling of the stress drop with magnitude (Cotton et al. 2008; Atkinson and Boore 2011). In the NGA models, authors proposed more complex magnitude scaling (e.g. bilinear, trilinear), but still these equations derived from larger-magnitude earthquake recording can overestimate the ground motions produced by smaller-magnitude events (Atkinson and Morrisson 2009). Bommer et al. (2007) conclude that for modeling the magnitude scaling over an extended range, this scaling could be linear at low magnitudes and then allow for a quadratic fall-off in slope over the upper range. They are currently working on it to understand which approach would be the best. Chiou et al. (2010) and Atkinson and Boore (2011) provided an update of their NGA models extended to lower magnitudes (down to 3.0) based on data from western

North America (median amplitudes are updated for 3 frequencies). It is not known yet if these modified models are predicting correctly low-magnitude ground-motions elsewhere.

**Purposes of the study**

In the short term, except from these two aforementioned models, reliable equations for predicting ground-motion amplitudes from very low to large magnitudes do not exist. Therefore, the purpose of the present study is to test the models selected in SHARE on the low-magnitude dataset available for the French accelerometric network. The first aim is to evaluate how these models perform on low-magnitude ground motions (M 4 to 5), in a magnitude range that is contributing in PSHA estimates obtained in this region. The second aim is to analyze the results in terms of regional variability of ground motions, keeping in mind that the interpretation will be limited as both aforementioned problems (namely magnitude scaling and regional variation) are playing a role and cannot be analyzed separately.

The first testing of GMPEs against weak motions recorded in active regions of Western Europe, using a reproducible technique, was proposed by Scherbaum et al. (2004) and applied in two following studies (Drouet et al., 2007 and Hintersberger et al. 2007). Scherbaum et al. (2004) proposed a likelihood-based technique to rank models according to their fit to observed data, through a categorization scheme. They illustrated the method on the records of a unique event, the Saint Dié earthquake (22/02/2003, $M_w$ 4.5 according to Drouet et al. 2010), at 13 rock stations. Hintersberger et al. (2007) performed an update of this study, using the same method applied on acceleration records from five earthquakes ($M_w$ 3.6 to 5.1) in the border region of Germany, France and Switzerland, resulting in a dataset made of 61 records with distances up to 300 km. Drouet et al. (2007) used 15 accelerometric records obtained in the Pyrenees from three earthquakes ($M_w$ 3.7, 3.7 and 3.9). In these studies, roughly the same set of GMPEs was tested, recordings at rock were considered and all frequencies available were mixed in the testing. The resulting best-fitting GMPEs were models derived from different tectonic context,

European, Western US, or Japanese. Therefore, these previous studies did not highlight regional variations of ground motions, but the results were interpreted with great caution, as the datasets were quite restricted. They posed the problem of testing GMPES derived from earthquakes with larger magnitudes than available in the datasets. The current study includes the dataset of Drouet et al. (2007), as well as part of the accelerometric data used in Scherbaum et al. (2004) and Hintersberger et al. (2007). GMPEs have evolved a lot, and in the following none of the GMPEs tested in these previous studies will be used as they all have been superseded.

The method chosen here for quantifying the goodness-of-fit of a GMPE to a dataset is the Scherbaum et al. (2009) technique (detailed in the section Method for testing GMPEs against observations). This method is new and has not been widely applied yet, hence we propose synthetic tests to evaluate the meaning of the "LLH" value reflecting the fit between a model and the data. Another issue which is not clear yet is the minimum number of data required for the Scherbaum et al. (2009) method to yield stable results. Applying the technique on a given dataset, we have no argument for stating that the results are independent on the sample, or can be considered as representative of the region in this magnitude range. Therefore, in the last part of the paper, we take advantage of a Japanese dataset to propose an answer to this question. As the Japanese dataset extends over a wide magnitude range, the question of the magnitude scaling is also addressed, by comparing results obtained from low and moderate-to-large magnitude datasets.

**Method for testing GMPEs against observations**

The recent method introduced by Scherbaum et al. (2009) is chosen for testing the models against the data. Scherbaum et al. (2009) provides a ranking criterion based on information theory (see the original paper for a detailed description of the theory behind). This technique is based on the probability for an observed ground motion to be realized under the hypothesis that a model is true. It provides one value, the negative average log-likelihood LLH (Delavaud et al. 2009), that reflects the fit between data and model:

$$LLH = -\frac{1}{N}\sum_{i=1}^{N} \log_2(g(x_i))$$

with N the number of observations $x_i$, and g the probability density function predicted by the GMPE (normal distribution). The ranking of models according to their fit to the data is then straightforward. In theory, it can be applied whatever the amounts of data available, but the results are expected to be more stable if the testing is performed on a large dataset. A small LLH indicates that the candidate model is close to the model that has generated the data, while a large LLH corresponds to a model that is less likely of having generated the data.

We propose to take advantage of synthetic data to understand the meaning of the LLH value (Fig. 1). In this respect, we step back from the information theory perspective, and simply concentrate on the calculated LLH values. Synthetic datasets are generated from an original Gaussian distribution, and distributions with modified characteristics (in terms of mean and sigma) are tested against these synthetic datasets. Results are displayed with increasing number of synthetic data (generated and tested), generating a new random dataset for each run so that the stability of the results can be verified. If testing the same distribution on the simulated dataset, mean LLH values obtained are close to 1.4-1.5. Then, if testing distributions that differ from the original one, mean LLH are increasing. For a distribution with a mean equal to the original mean plus one sigma, or a sigma twice the original sigma, LLH values are around 2.0 (both distributions are roughly providing the same fit to the data). In the worst case considered in this example, the tested distribution has a mean equal to the original mean plus 2.5 sigma, and a sigma equal to 0.8 times the original sigma, producing mean LLH values as high as 9-10. These synthetic tests give us an idea of the LLH value to expect depending on the distribution of the observations with respect to the model. Besides, these simulations also yield a rough idea about the minimum number of observations required for obtaining a stable mean LLH. Based on the results in Figure 1, mean LLH values are reached from ~40 observations on. However, here the synthetic dataset is perfectly distributed according to a normal distribution, which is not true in the real cases. Hence, 40 observations must be considered as a minimum, and more tests must be performed on different observation datasets to clearly define this minimum number of data.

**GMPEs best adapted to the French weak motion data**

*Description of the data*

The accelerometric French network RAP (Fig. 2) has been operating since roughly 15 years, with more and more stations installed since 1996 (Réseau Accélérométrique Permanent, Péquegnat et al. 2008). Only earthquakes recorded in metropolitan France will be taken into account here (a separate study would be required for earthquakes belonging to the tectonic context of the French Antilles). The RAP stations are dial-up or continuous recording stations. They consist of one 3C broadband accelerometric sensor (Kinemetrics episensors, except for some of the oldest stations having Guralp CMG5). They are connected to a 24-bit three-component digitizer sampling at 125 Hz. The full scale of the channel corresponds to ±1g for all the stations used in this paper. The useful frequency band is 0-50 Hz. Only offset correction is applied to the data without any additional filtering. The data used in this paper has been visually cleaned by checking the signal-to-noise ratio and the time accuracy on the three components.

Using only high quality accelerograms, a total of 16 events with moment magnitudes between 3.8 and 4.5 are available (Fig. 2 and 3, Table 1). These earthquakes belong to the seismically active parts of France (Pyrenees, Alps and Lower Rhine Embayment), which have been classified as "active shallow crustal regions" in Delavaud, Cotton et al. (2012). Most of the GMPEs will be tested outside their validity range, but to limit the extrapolation below their minimum magnitude bound, moment magnitudes considered here are higher or equal to $M_w$=3.8. Only a selection of the available Mw3.8 events is included in this study (the events with the greatest number of recordings). Thus, GMPEs are tested against a dataset that is roughly homogeneously distributed in terms of magnitude, and the results are not influenced by the presence of more low magnitudes than larger ones (Fig. 3). The level of knowledge on the site conditions in the French accelerometric network varies greatly from one station to the other. Sites are classified according to the four ground categories defined in the current European seismic code EC8 (CEN 2004). At some sites, some geophysical and geological studies have been led. For the sites with a "medium to high" confidence level in the estimation of the shear-wave velocity in the top 30 m, the estimated Vs30 has been used. Elsewhere, for the NGA models relying on the Vs30, a mean Vs30 value corresponding to the EC8 class has been used (1000 m/s for A, 600 for B, 250 for C, 100 for D). We refer the reader to the Regnier et al. (2010) report for a detailed study on the site conditions of the accelerometric stations. Considering source-site distances up to 300km, a total of 191 recordings is obtained (Table 1). As the considered magnitudes are small, the size and extension of the fault planes have negligible impact on the calculation of the different distance measures. Moment magnitude is available for 12 out of 16 events (Drouet et al., 2010). For the 4 remaining events, a magnitude correlation is applied (Drouet et al. 2010). Focal mechanism is known for 12 out of the 16 events (Table 1). However the influence of the style of faulting is unlikely to be significant for small events that can be approximated by point sources for most recording locations.

*Description of the GMPEs considered*

Several recent models have been selected for testing the accelerometric dataset (Table 2, Fig. 4), although more than 180 equations are listed in Douglas (2011) for elastic response spectral ordinates. This short list consists in equations developed for active shallow crustal regions, in terms of moment magnitude, and including a non-linear magnitude scaling term (except for Cauzzi and Faccioli 2008). The models Akkar and Bommer (2010), Cauzzi and Faccioli (2008), Zhao et al. (2006), and Chiou and Youngs (2008), have been selected within the SHARE project for application in shallow crustal regions (see Delavaud, Cotton et al. 2012, for a detailed description of the selection process). Here two more NGA models often considered in current engineering seismology projects are tested, namely Abrahamson and Silva (2008) and Boore and Atkinson (2008), as well as the new Chiou et al. (2010) equation extended to lower magnitudes. Furthermore, the Bindi et al. (2009) model, that showed to predict well the SHARE strong motion dataset (Delavaud, Cotton, et al. 2012), is also considered.

*Akkar and Bommer (2010)* have developed a pan-European equation, predicting geometrical mean of horizontal pseudo-spectral accelerations for magnitudes ranging from 5 to 7.6, at distances up to 100km (Joyner and Boore distance). The spectral period range is 0.05-3 seconds. The generating dataset is covering several countries from Europe and the Middle East, from moderate to high seismicity.

*Cauzzi and Faccioli (2008)* model predicts geometrical mean of accelerations for magnitudes ranging from 5 to 7.2, at distances up to 150km (hypocentral distance). The spectral period range is 0.05 - 20 seconds. Such long periods are crucial for prediction of ground motions for bridges and tall buildings. This equation has been initially developed for application in Italy but it is based on a worldwide crustal earthquake dataset. A large part of this dataset (~80%) is coming from the Japanese K-NET strong motion network (see Section Data and Resources), 5% is coming from Europe and Turkey. The model handles two definitions for the site conditions, either directly using the Vs30 as predictor variable, or using the Eurocode 8 ground categories (CEN 2004). This model has one limitation as it is defined for hypocentral distances larger than 15km.

*NGA models (2008)* have been developed from a worldwide dataset (including events from the Euro-Mediterranean region) for predicting ground motions in the Western United States, at distances up to 200km, and for spectral periods ranging from 0.01 to 10 seconds. Analytical models based on numerical simulations are included, providing constraints on the ground-motion scaling outside the range well constrained by the empirical data. Here, three of these models are considered. *Abrahamson and Silva (2008)* and *Boore and Atkinson (2008)* equations apply for magnitudes higher or equal to 5. *Chiou and Youngs (2008)* model is in theory applicable for magnitudes higher or equal to 4. These NGA models require some predictor variables that are not known for the French accelerometric stations and that must be estimated: depth-to-top of rupture, and depth to the 1 km/s shear-wave velocity horizons. The Boore and Atkinson (2008) model uses the smallest number of predictor variables of the NGA models. All NGA models predict site response on the basis of the average shear-wave velocity in the top 30 m.

*The model of Chiou et al. (2010)* has been developed for comparing weak ground motions between California and other active tectonic regions. For now, coefficients are available for 3 periods (PGA, 1 and 0.3 sec). The equation is developed for small-to-moderate shallow crustal earthquakes ($3<M<5.5$) up to 200 km distance, and has been derived from Californian data. The specific goal of the authors is "to provide an empirical model that can be confidently used in the investigation of ground-motion difference between California and other active tectonic regions […] where the bulk of ground-motion data from shallow crustal earthquakes is in the small-to-moderate magnitude range" (p. 1 of Chiou et al., 2010). Both the equations for southern California and central California will be tested.

*Bindi et al. (2009)* is an equation derived from Italian data only. The generating dataset is made of magnitudes from 4.0 to 6.9 recorded at distances up to 100km. The spectral period range is 0.03-2 seconds.

*Zhao et al. (2006)* is aimed at predicting ground motions in Japan. The dataset contains distances up to 300km and magnitudes between 5 and 7.3 (crustal earthquakes). The spectral period range is 0.05 - 5 seconds. Most of the data have been recorded in Japan, except for a few overseas events providing short source distance recordings.

*Parameter compatibility issue*

All GMPEs considered in this study use the moment magnitude scale to characterize earthquake size. The distance measure is different from one model to the other, e.g. some models use the Joyner and Boore distance measure (which is measured horizontally on the surface, e.g. Boore and Atkinson 2008) while others are based on the rupture distance (closest distance to the rupture plane). Each model is applied with its native distance measure. The Beyer and Bommer (2006) conversions are used to take into account different definitions in the horizontal component (geometrical mean, etc.). The NGA models predict site response on the basis of the average shear-wave velocity over the upper 30m (Vs30), whereas European equations take into account three generic site classes: rock, stiff soil and soft soil (corresponding to shear-wave velocity intervals, CEN 2004).

*Results*

The equations are tested against the homogeneous dataset described above. Although most models have been developed for maximum distances varying from 100km to 200km (Table 2), distances as far as 300km are taken into account to ensure a minimum number of data (191 recordings). In a second step, the same calculations are performed on a reduced dataset selecting distances up to 200km (143 recordings), to check that the results remain stable. As shown in Figure 4, all LLH values are roughly between 2.5 and 4.5. The synthetic tests showed that a "perfect" fit would yield LLH=1.4-1.5 (Fig. 1, section Method for testing GMPEs against observations). Three models are yielding the lowest and most stable LLH values over the whole frequency range: Cauzzi and Faccioli (2008, CF2008), Akkar and Bommer (2010, AB2010) and Abrahamson and Silva (2008, AS2008). These three equations result in LLH values varying from 2.5 up to 3.5. The Zhao et al. (2006) GMPE is not included in this best-fitting GMPE short-list; compared to the three above mentioned models, this GMPE is providing slightly higher LLH on average over the whole frequency range. The Abrahamson and Silva (2008) model requires some parameters describing the site and the source, which are not well constrained (depth-to-top of rupture, fault dip, downdip rupture width) or not available (the depth to Vs=1000 m/s). We performed sensitivity studies on these parameters, and we observed that, if using a reasonable (but still wide) range of values, there is very little impact (if any) on the LLH obtained. This might be due to the low magnitudes involved and the source-site distances available. Sensitivity tests performed for the Chiou and Youngs (2008) NGA model, and the Chiou et al. (2010), led to the same conclusions.

Some models show a good ability to predict the observations only for part of the frequency range. The model of Chiou and Youngs (2008, CY2008) performs roughly well only for low frequencies (<3Hz). Conversely, the model of Bindi et al. (2009, B2009) predicts observations correctly only for higher frequencies, larger than ~3Hz. These results highlight the need to test GMPEs as a function of spectral frequency. If mixing frequencies, a mean LLH value would be obtained which would not be reflecting correctly the goodness-of-fit of the model to the data. Delavaud et al. (2009) also observed such a strong dependence with frequency while applying the LLH-based method on Californian data. It is interesting to observe that, if considering the results at 1 Hz, except for the Bindi et al. (2009) model, the LLH are all concentrated in a narrow interval (2.9-3.3). For this frequency, models' performances are comparable.

The Chiou et al. (2010) model, derived specifically for predicting ground motions for magnitudes lower than 5.5, consists in two sets of coefficients respectively for Central and Southern California. The LLH values at 0.3 seconds and PGA are very high using the coefficients for Central California (6.5 at 0.3 sec, and ~5 at the PGA). The equation for Southern California fits well the data at 1 Hz and at the PGA, but is yielding a higher LLH value than the rest of the models for the period 0.3s. Low-magnitude ground motions in California might not be similar to low-magnitude ground motions in our target region. However, this result must be taken with caution, as only 3 frequencies are available for comparison (Chiou et al. 2010).

Furthermore, to visualize the fit between the data and the predictions, a more classical technique is to display the residuals. The residual is the difference between the prediction and the observation in terms of logarithm, normalized by the sigma of the model. Some residuals are displayed in Figs. 5 to 8 to illustrate the fit for two of the best-fitting model (CF2008, Fig. 5, and AB2010, Fig. 8), and for two of the models predicting higher LLH values (AB2008, Fig. 6, Z2006, Fig. 7). These histograms provide complementary insights on the fit between models and data. They show that the observations are characterized by a higher variability than the predicted distributions, which is expected as motions from small earthquakes have proved to be more variable than motions from larger earthquakes (e.g. Youngs et al. 1995). The origin of this aleatory variability is not identified yet (either a true physically-based uncertainty or an uncertainty due to metadata, Bommer et al. 2007). Besides, whenever the median of observations does not fit the median of predictions, the models are over-predicting the amplitudes (e.g. Fig. 6 displaying the results for the Boore and Atkinson 2008 model). This observation is also expected from different past studies (e.g. Bommer et al. 2007). Residual histograms are shown for varying maximum distances (300, 200 and 100km). If reducing the maximum distance to 200km, or considering only rock stations (and thus reducing the uncertainty on the site conditions), the ranking obtained for GMPEs remains stable. For 100km, we believe that there is too few data to derive reliable conclusions.

The equations best fitting the French accelerometric weak-motion dataset have been highlighted. Two models selected within SHARE for crustal regions are fitting reasonably well the data (CF2008, AB2010). No significant regional variation of ground motions is highlighted, and the magnitude scaling could be predominant in the control of ground-motion amplitudes. However, clear explanations for the relative good performance of these models are not straightforward. These GMPEs are imported models and they are applied at magnitudes lower than their minimum magnitude validity limits. The two other models selected in SHARE, Boore and Atkinson (2008) and Zhao et al. (2006) are slightly over-estimating the data, which is coherent with many recent studies (e.g. Cotton et al. 2008). One question that is naturally raised is whether the ranking deduced from these low-magnitude motions would hold if moderate magnitudes were available. However, at this stage, we have no argument to assert that if available, stronger ground motion would also match these GMPEs. In the following, we make use of a Japanese dataset to tackle some of the unresolved questions that appeared during the testing on the weak-motion accelerometric dataset.

**Testing predictions and observations on the Japanese data**

The rich Japanese dataset contains both weak and strong ground motions, corresponding to a wide magnitude range. The KiK-net and K-NET networks recordings have been collected up to the end of 2009 (Laurendeau et al. 2011). Only events characterized in the F-net catalog are selected in order to have consistent meta-parameters (Mw, hypocenter location, focal depth and rake angle). Besides the data subset used in this study includes only crustal events (focal depth ≤25km and excluding offshore events on the subduction side) and rock sites (VS30≥ 500 m/s, with $V_{S30}$ deduced from KiK-net velocity profiles, Boore et al., 2011). A magnitude-distance filter was applied according to Kanno et al. (2006) predictions, taking 2.5 gal as a PGA threshold. S-wave triggered and multi-events records have been eliminated. The source distance is the hypocentral distance for events with $M_w$ < 5.7 and the closest distance from the fault plane to the observation site for the events with larger magnitudes.

A set of eight models is selected, including recent models derived from Japanese data or from other active crustal regions of the world. At first, the aim is to test a dataset with characteristics close to the French accelerometric dataset to determine if the number of recordings available is sufficient for considering the results reliable. Secondly, the same GMPEs are tested against the Japanese dataset on the moderate-to-large magnitude range. The objective is to analyze the performance of the models according to the magnitude range considered. In other words, using the Scherbaum et al. (2009) method, the ranking of the models obtained on the low magnitude range is compared with the ranking of the models resulting from the larger magnitude range.

At first, only recordings corresponding to earthquakes with magnitudes between 4 and 4.9 are considered (with at least 8 recordings per event). The difference between this dataset and the French dataset is in the distance distribution, the Japanese network is much denser and distances available in our database are shorter. For each run (each curve in Fig. 9), subsets are extracted at random from the initial dataset with a condition on the total number of recordings, the sample must be constituted of 170 to 210 records (so 190 in average). The resulting sample is made of 11 to 18 earthquakes, distributed all over Japan. An example of a random dataset is displayed in Figure 10. The range of LLH values obtained (up to 5.2) is comparable to the LLH obtained on the French weak motion data. Three models provide low LLH, between 1.5 and 2.5, implying a good fit with the data (Kanno et al. 2006, Cauzzi and Faccioli 2008, Chiou and Youngs 2008). The results confirm again that the fit between the predictions and the observations varies with the frequency considered. For frequencies higher than 6.0, all tested models are providing stable LLH values over the frequency range, restricted to a narrow interval (LLH=1.8-2.8). Most important for this test of stability, the results do not change much from one subset to the other, which implies that the rough hierarchy between models can be obtained with a small dataset of recordings. This dataset has the same size as the French weak motion dataset tested, but with a wider distance range in the case of the French dataset. The models Cauzzi and Faccioli (2008), Kanno et al. (2006), and Chiou and Youngs (2008), are identified as the best-fitting models, with the lowest LLH over the whole frequency range. The model with the poorest fit is Atkinson and Boore (2008). GMPEs are then grouped according to their ranking (Table 3).

Next, all recordings corresponding to earthquakes with magnitudes between 5 and 7 are considered (Fig. 11, around 1200 recordings if considering events with at least 10 recordings). Again, LLH values obtained for the different GMPEs range in a rather narrow interval for frequencies 6 to 10Hz (1.9-2.4), but show to be quite different for frequencies lower than 5-6 Hz (1.4-3.2). Three models emerge as the best-fitting equations over 0-10Hz: Cotton et al. (2008), Cauzzi and Faccioli (2008) and Kanno et al. (2006), with LLH values varying from 1.5 to 2 maximum. It is worth noting that the models Cotton et al. (2008) and Kanno et al. (2006) have been derived from a database of Japanese recordings, whereas the model Cauzzi and Faccioli (2008) is based on a database made of ~80% Japanese recordings. These results would tend to support the idea that ground motions in Japan are displaying specific features

(regional specificity). Some similar findings were obtained by Delavaud, Scherbaum et al. (2012) based on two frequencies (1Hz and PGA). Two other equations are yielding rather stable LLH values on the whole frequency range, Zhao et al. (2006) and Chiou and Youngs (2008), however with slightly higher LLH values in the lower frequency range (<4Hz, 1.8<LLH<2.4).

Based on the hierarchy obtained from the LLH values, the models are ranked in 4 categories (Table 3), from the best-fitting models (1.5<LLH<1.8 for f<6Hz) to the worse (1.7<LLH<3.4 for f<6Hz). Comparing the results of the testing for low and larger magnitudes, it is interesting to observe that the hierarchy between models is only partly preserved. For two models the ranking obtained from the low magnitude dataset is rather different than the ranking obtained from the larger magnitude dataset: Cotton et al. (2008) model is no longer ranked among best-fitting models (LLH around 2.5), whereas Chiou and Youngs (2008) is now among the best-fitting models (LLH around 2.0). Moreover, examples of normalized residuals are displayed in the case of the Cauzzi and Faccioli (2008) model (Fig. 12). The histograms highlight the link between a low LLH (~1.6) and the good fit of the normalized residual distribution with respect to the standard normal distribution.

**Conclusions**

We have analyzed and quantified the coherency between several GMPEs and three datasets, a low magnitude (3.8-4.5) dataset of recordings from the French accelerometric network, and two datasets build from the Japanese K-NET and KiK-net networks. From these studies we derive several conclusions and highlight remaining key questions.

The Scherbaum et al. (2009) technique, relying on the calculation of a loglikelihood LLH, is a very practical and powerful tool to quantify the fit between predictive equations and observations. We find that for LLH values reaching 1.5-1.6, the distribution of the normalized residuals is matching well a standard normal distribution; whereas for values higher than ~3-4, either the mean, the sigma, or both values calculated from the residual distribution are strongly moving away from the parameters of the standard normal distribution.

The fit between observations and predictions proved in several cases to vary greatly with the frequency. When enough data is available, the testing and application of the Scherbaum et al. (2009) technique (or any other technique for testing GMPEs against data) should be carried out separately for each frequency. Otherwise some information is lost, and a mean LLH is calculated, which might not represent well individual LLH per frequency.

The analysis of the dataset from the French accelerometric network brings new insights for low-to-moderate seismicity regions of Western Europe (shallow active regions). The three models yielding lowest LLH values on the French accelerometric low-magnitude dataset over the whole frequency range are the Cauzzi and Faccioli (2008), Akkar and Bommer (2010) and Abrahamson and Silva (2008) equations. These models, derived from different crustal tectonic environments, are thus the equations most coherent with the weak motions dataset recorded in active regions of France (Alps, Pyrenees and Lower Rhine Embayment). Both models CF2008 and AB2010 have been selected in SHARE for application in active shallow crustal regions across Europe. Akkar and Bommer (2010) is a pan-European model, whereas Cauzzi and Faccioli (2008) is mostly relying on Japanese data, and Abrahamson and Silva (2008) on California and world-wide data. This result does not highlight regional variation of ground motions.

The sensitivity studies carried out on the Japanese database show that considering a subset with properties similar to the French accelerometric dataset (in terms of magnitudes and amount of recordings), the ranking of GMPEs obtained does not depend on the subset. This finding implies that the results obtained from the French accelerometric dataset, made of 191 recordings at distances less than 300km, can be considered as stable (until more data is available to prove it, especially more short distance recordings).

The same set of predictive models is tested on the moderate-to-large magnitude dataset from Japan (5≤M≤7). Comparing the ranking of GMPEs obtained from this larger magnitude range with the ranking resulting from the low magnitudes, some features are maintained, but for some models the ranking is modified. The magnitude scaling is therefore controlling the ground motions for some of the GMPEs tested. An interesting observation is that, when considering magnitudes that fall in the validity limits of the equations, the models predicting the best the observations are all native Japanese models (with LLH values of 1.5-1.7 indicating that the fit is very good). Ground-motion scaling in Japan might differ significantly from other active regions.

## Data and Resources

All the accelerometric data from the French Accelerometric network is publicly available online at http://www-rap.obs.ujf-grenoble.fr/ (last accessed April 2012), as well as the accelerometric data from Japan (www.k-net.bosai.go.jp, last accessed April 2012), and the earthquake data from the Réseau National de Surveillance Sismique (RéNaSS, http://renass.u-strasbg.fr/, last accessed April 2012). The SHARE project is presented here: http://www.share-eu.org/ (last accessed April 2012).

## Acknowledgments

Most of this work has been funded by the European Research program FP7, entitled 'Seismic Hazard Harmonization in Europe' (SHARE, contract number 226967). H. Oksuz-Tasan benefited from a scholarship from the MEEES international master program (www.meees.org/). We are indebted to the National Research Institute for Earth Science and Disaster Prevention (NIED), Japan, and to the French Accelerometric Network (RAP) for providing the data for this analysis. Finally, comments by S. Drouet and E. Faccioli contributed to improve significantly the clarity of the manuscript.

in Japan using site classifications based on predominant period, *Bull. Seismol. Soc. Am*. 96, 3, 898–913.

ISTerre

IRD-UJF-CNRS-IFSTTAR

BP 53

38041 Grenoble Cedex 9, France

C.B., H.T., A.L., F.C., P.G.

Swiss Seismological Service,

Institute of Geophysics, ETH

Zurich, Sonneggstrasse 5, NO, 8092 Zurich, Switzerland

E.D.

Institute of Earth and Environmental Sciences,

University of Potsdam,

Karl-Liebknecht-Strasse 24-25, 14476 Golm, Germany

N.K.

# Tables

Table 1. Description of the earthquakes and corresponding recordings used in the study. Magnitudes have all been calculated by Drouet et al. (2011), except when specified.

| Event dates | $M_w$ | Meca. (‡) | Long. (°) | Lat. (°) | Prof. (km) | Number of stations (≤300km and ≤200km) | | Reference Coordinates | Reference Mechanism |
|---|---|---|---|---|---|---|---|---|---|
| 31-10-1997 | 4 | U | 6.57 | 44.26 | 2 | 11 | 10 | Drouet et al. 2010 | No reference |
| 21-08-2000 | 4.4(*) | U | 8.44 | 44.86 | 10 | 11 | 11 | RéNaSS | No reference |
| 25-02-2001 | 4.5(§) | R | 7.47 | 43.49 | 14 | 5 | 5 | BCSF (2001), Bureau Central Sismologique Français | |
| 16-05-2002 | 4 | N | -0.143 | 42.922 | 9.5 | 9 | 9 | Drouet et al. 2010 | Chevrot et al. (2011) |
| 11-12-2002 | 3.8 | N | -0.33 | 43.04 | 5 | 5 | 5 | Drouet et al. 2010 | Chevrot et al. (2011) |
| 12-12-2002 | 4 | N | -0.28 | 43.11 | 10 | 9 | 7 | Drouet et al. 2010 | Chevrot et al. (2011) |
| 21-01-2003 | 3.8 | N | -0.36 | 43.05 | 10 | 12 | 7 | Drouet et al. 2010 | Chevrot et al. (2011) |
| 22-02-2003 | 4.5 | N | 6.66 | 48.34 | 10 | 13 | 9 | Drouet et al. 2010 | BCSF (2003) |
| 11-04-2003 | 4.3(*) | U | 8.97 | 44.81 | 5 | 21 | 10 | RéNaSS | No reference |
| 23-02-2004 | 4.2 | SS | 6.28 | 47.3 | 10 | 19 | 17 | Drouet et al. 2010 | BCSF (2004) |
| 18-09-2004 | 4.6 | N | -1.6 | 42.78 | 2 | 9 | 6 | Drouet et al. 2010 | Chevrot et al. (2011) |
| 30-09-2004 | 4.1 | N | -1.45 | 42.77 | 10 | 8 | 6 | Drouet et al. 2010 | Chevrot et al. (2011) |
| 08-09-2005 | 4.4 | SS | 6.87 | 46.01 | 10 | 22 | 12 | Drouet et al. 2010 | RAP (2005) |
| 17-11-2006 | 4.5 | N | 0.01 | 43.08 | 9.7 | 18 | 15 | Drouet et al. 2010 | Sylvander et al. (2008) |
| 30-07-2007 | 4.0(*) | U | 9.71 | 44.92 | 10 | 5 | 0 | RéNaSS | No reference |
| 15-11-2007 | 4.0(*) | N | 0.0 | 43.01 | 8 | 14 | 14 | BCSF (2008) | |

(*) Calculated from the RéNaSS local magnitude using the Drouet et al. (2010) Mw-ML_RéNasSS correlation (RéNaSS stands for Réseau National de Surveillance Sismique).

(§) Moment magnitude from ETH-SED

(‡)SS, strike slip, N normal, R reverse, U unknown

**Table 2: Ground-motion prediction equations used in the study**

| GMPE reference | GMPE Acronym | Magnitude validity bounds | Frequency range (Hz) | Max. source-site distance (km) | Region of the generating dataset |
|---|---|---|---|---|---|
| Bindi et al. 2009 | B2009 | 4.0-6.9 | 0.5-33.33 | 100 ($R_{JB}$) | Italy |
| Cauzzi & Faccioli 2008 | CF2008 | 5.0-7.2 | 0.05-20.0 | 150 ($R_{RHYP}$) | K-Net+ worldwide |
| Kanno et al. 2006 | Ketal06 | 5.2-8.2 | 0.2-20.0 | 300 ($R_{RUP}$) | Japan (depth<30km) |
| Chiou & Youngs 2008 | CY2008 | 4.0-8.0 | 0.1-100.0 | 200 ($R_{RUP}$) | worldwide |
| Chiou et al. 2010 (central) | Ccentral | 3.0-5.5 | PGA, 3.3, 1 | 200 ($R_{RUP}$) | Central California |
| Chiou et al. 2010 (southern) | Csouth | 3.0-5.5 | PGA, 3.3, 1 | 200 ($R_{RUP}$) | Southern California |
| Cotton et al. 2008 | Cetal08 | 4.1-7.3 | 0.3-100.0 | 100 ($R_{RUP}$) | Japan |
| Zhao et al. 2006 | Z2006 | 5.0-7.3 | 0.2-20.0 | 300 ($R_{RUP}$) | Japan |
| Boore and Atkinson 2008 | BA2008 | 5.0-8.0 | 0.1-100.0 | 200 ($R_{JB}$) | worldwide |
| Abrahamson & Silva 2008 | AS2008 | 5.0-8.5 | 0.1-100.0 | 200 ($R_{RUP}$) | worldwide |
| Akkar & Bommer 2010 | AB2010 | 5.0-7.6 | 0.33-20.0 | 100 ($R_{JB}$) | Europe + Middle East |

**Table 3: Results of the testing on the Japanese dataset: classification of the GMPEs according to their fit to the data. Two data sets have been considered separately (see the text).**

| Ranking according to LLH: | Larger magnitude range | Low magnitude range |
|---|---|---|
| Best fitting models | CF2008, Ketal06, Cetal08 | CF2008, Ketal06, CY2008 |
| Intermediate | Zetal06, CY2008 | Cetal08, AB2010 |
| Poorly fitting models | AB2010, AS2008 | AS2008, Zetal06 |
| Worse fit | AB2008 | AB2008 |

# Figure Captions

Figure 1: Synthetic data simulation to evaluate the LLH values meaning. Left hand corner: the original distribution (black) and the distributions that are tested. The original distribution ($\mu_i$, $\sigma_i$) corresponds to the ground motion predicted by Akkar and Bommer (2010) model for a magnitude $M_w$=4 at $R_{jb}$=40km (natural logarithm of PGA in m.s$^{-2}$). In the 5 graphs, a candidate distribution is tested against a dataset generated from the original distribution. The mean and sigma of the candidate distribution is indicated in the title of each graph. The synthetic dataset is increased step by step (from 1 to 221 samples, the whole sample is randomly generated at each step). Individual LLH (gray points) and mean LLH (solid line) are computed for each run.

Figure 2: Location of earthquakes considered in the study (stars, Table 1) and stations of the French Accelerometric Network RAP (triangles).

Figure 3: Distribution of the data used in this study, in terms of distance source-to-site and magnitude $M_w$ (191 recordings in total, see Table 1).

Figure 4: Quantifying the fit between observed spectral accelerations (French accelerometric data, Fig. 2) and corresponding predictions provided by a list of GMPEs, LLH value versus frequency considered. See Table 2 for the GMPEs acronyms. Note that Chiou et al. (Ccentral/Csouth) is defined only for 3 frequencies (1, 3.33Hz, PGA), the results at PGA have been positioned at 50Hz for graphical reasons.

Figure 5: Histogram of residuals superimposed to the standard normal distribution representing the Cauzzi and Faccioli (2008) model, using the French accelerometric subset described in the text, at 3.3 Hz. A residual z corresponds to [Log(observation)-Log(prediction)]/sigma. The Gaussian with mean and sigma calculated from the residuals is superimposed to the histogram. From left to right: maximum distance considered is successively 300, 200 and 100 km (corresponding to LLH=2.58, 2.60, 2.18). The number of data is decreasing accordingly: 191, 143, 75 recordings.

Figure 6: See legend of Fig. 5. In this case the model tested is Boore and Atkinson (2008), at 2 Hz. From left to right: maximum distance considered is successively 300, 200 and 100 km (corresponding to LLH=4.13, 4.48, 5.35).

Figure 7 : See legend of Fig. 5. In this case the model tested is Zhao et al. (2006), at 2 Hz. From left to right: maximum distance considered is successively 300, 200 and 100 km (corresponding to LLH=3.4, 3.68, 4.3).

Figure 8 : See legend of Fig. 5. In this case the model tested is Akkar and Bommer (2010), at 2.3 Hz. From left to right: maximum distance considered is successively 300, 200 and 100 km (corresponding to LLH=2.68, 2.8, 2.76).

Figure 9: Testing GMPEs against the low-magnitude Japanese dataset (M4-4.9): loglikelihood LLH values obtained on the Japanese dataset, versus frequency. Five subsets are considered for each GMPE. Each subset is randomly extracted from the original dataset (condition: 170 to 210 recordings, resulting in 11 to 18 earthquakes).

Figure 10 : Example of a low-magnitude dataset randomly extracted from the original Japanese dataset (185 recordings and 11 events).

Figure 11: Testing GMPEs against the moderate-to-large magnitude Japanese dataset (M5-7): loglikelihood LLH values, versus frequency. See Table 2 for the GMPEs acronyms.

Figure 12: Predictions from the Cauzzi and Faccioli (2008) model compared to the Japanese accelerometric subsets (at 1.25Hz): histogram of normalized residuals superimposed to the standard normal distribution representing the model. Left: subset containing 1143 recordings corresponding to 36 events with $5 \leq Mw \leq 7$ (LLH=1.65). Right: subset containing 185 recordings corresponding to 15 events with $4 \leq Mw \leq 4.9$ (LLH=1.6). The Gaussian with mean and sigma calculated from the residuals is superimposed to the histogram.

# Figures

## Fig. 1

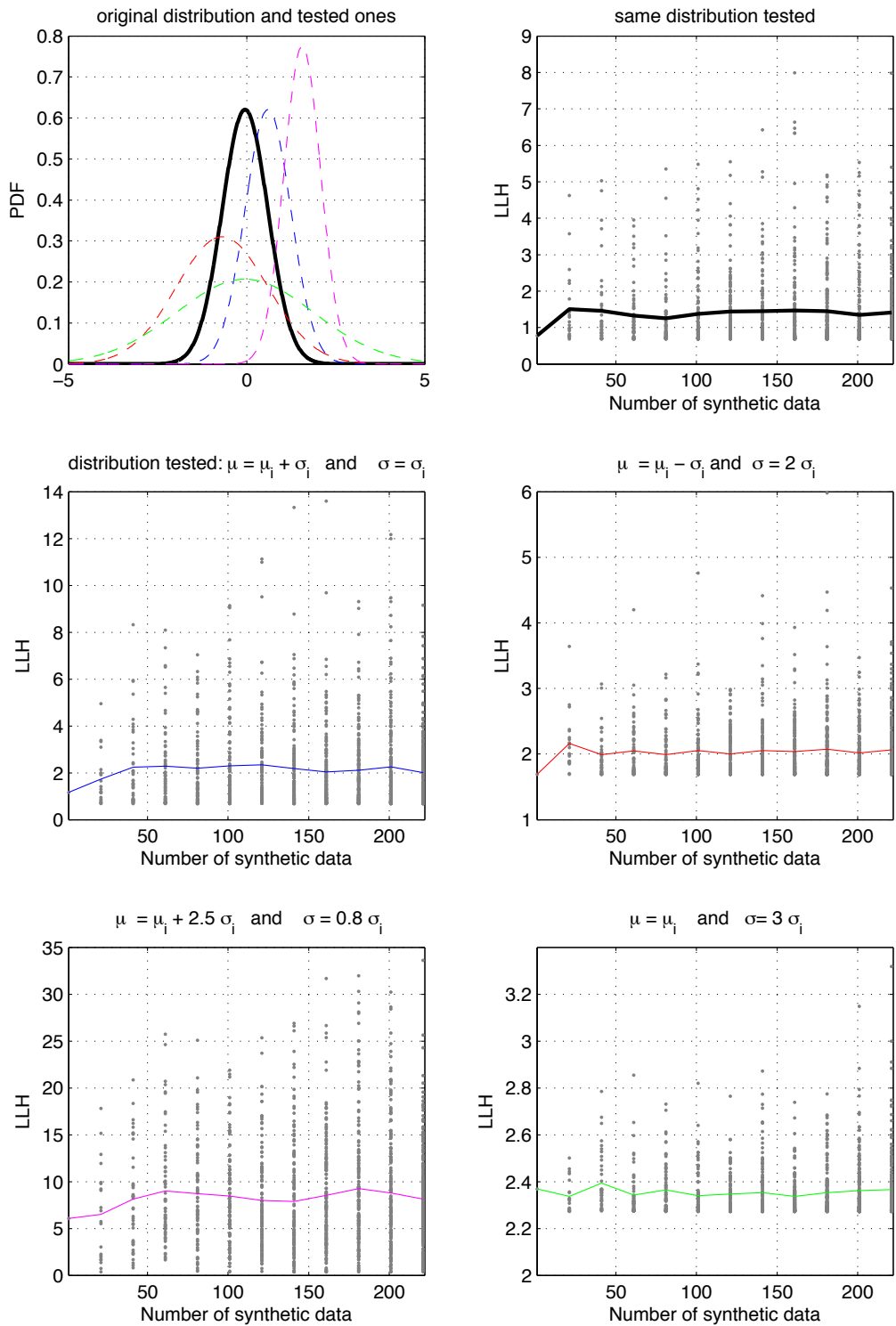

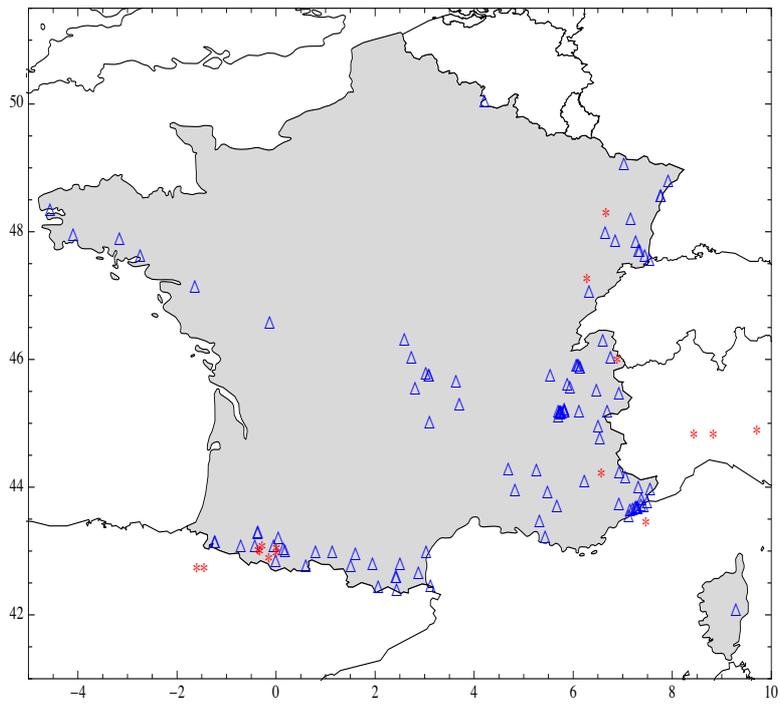

**Fig. 2**

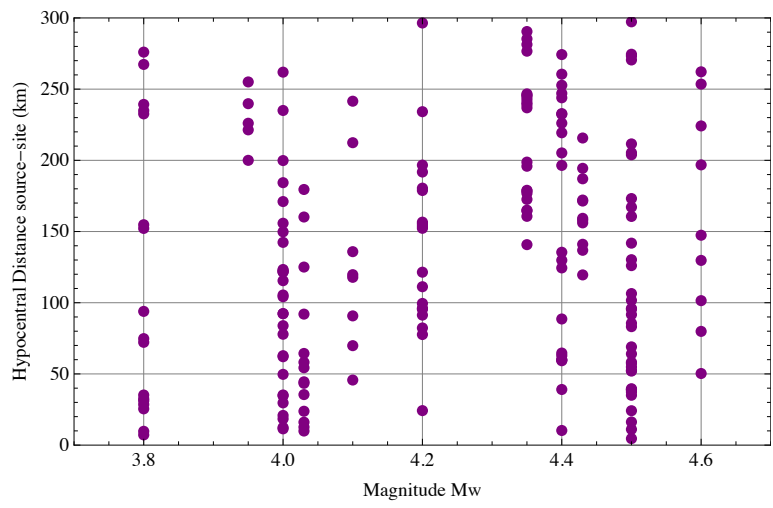

**Fig. 3**

**Fig. 4**

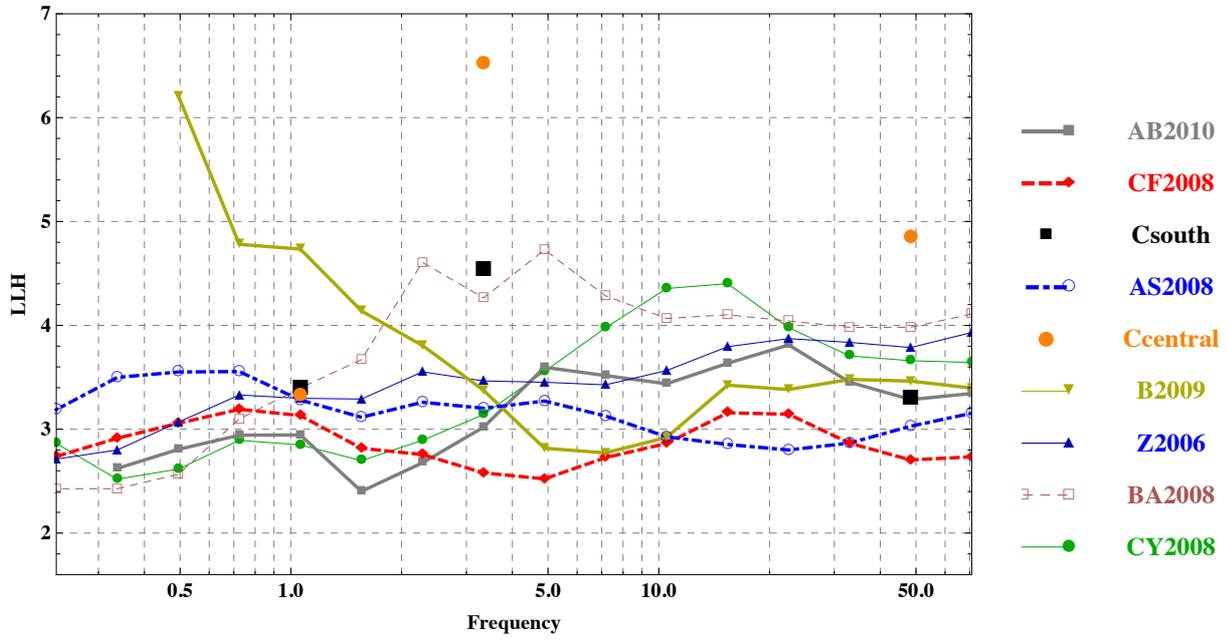

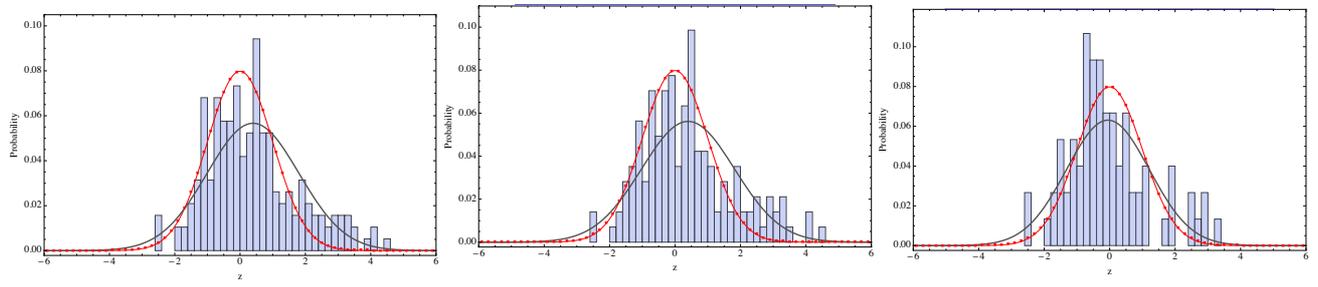

**Fig. 5**

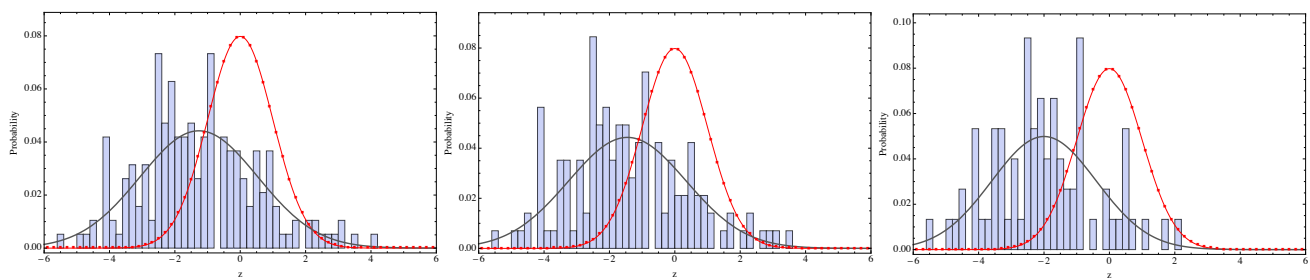

**Fig. 6**

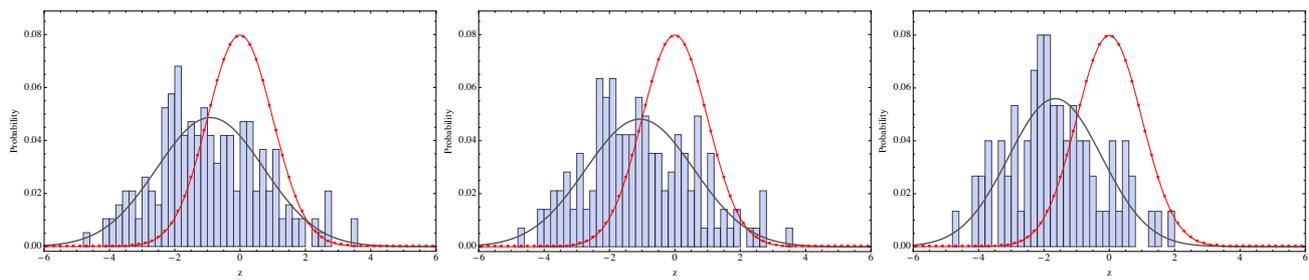

**Fig. 7**

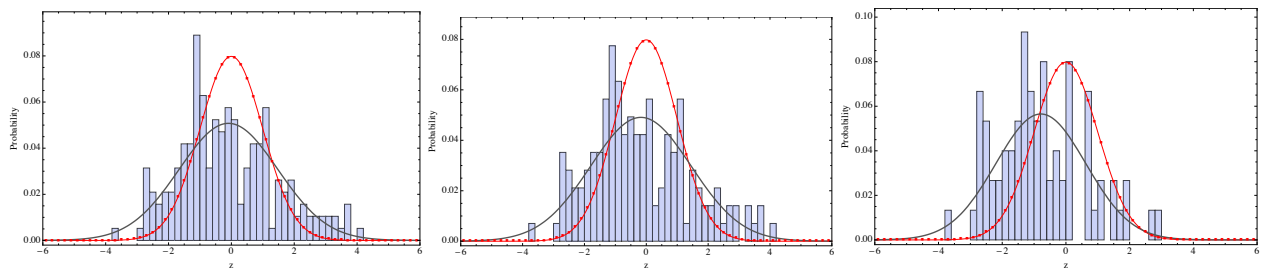

**Fig. 8**

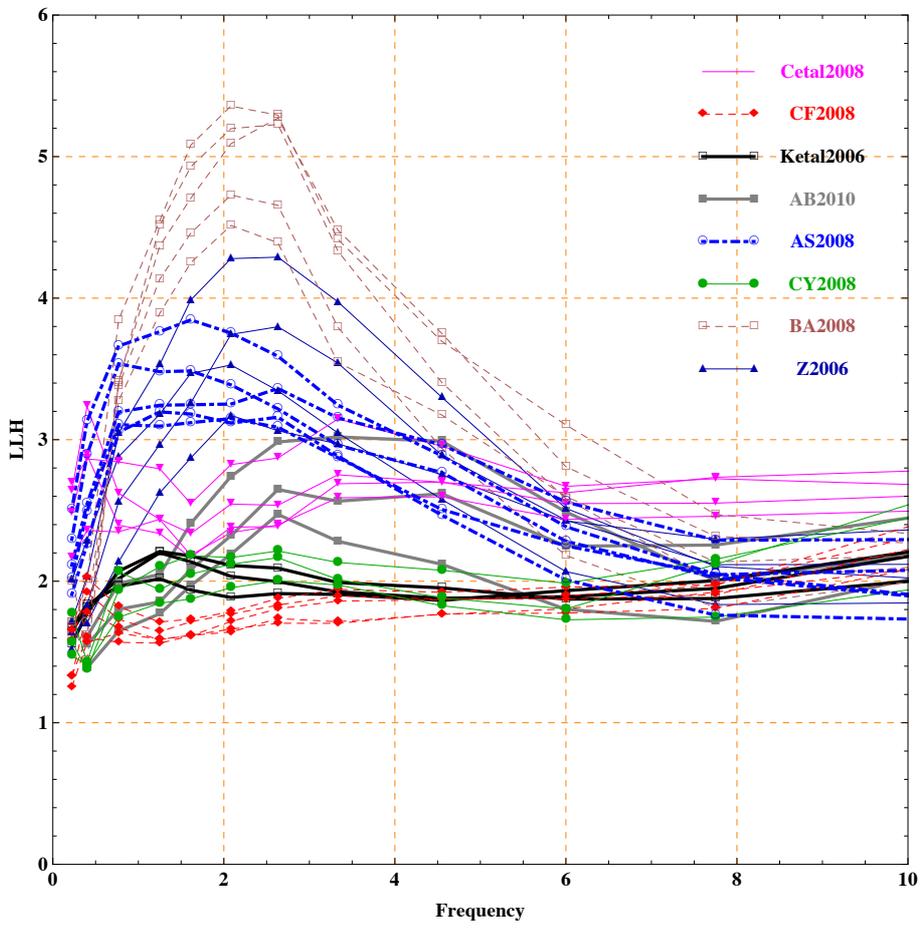

**Fig. 9**

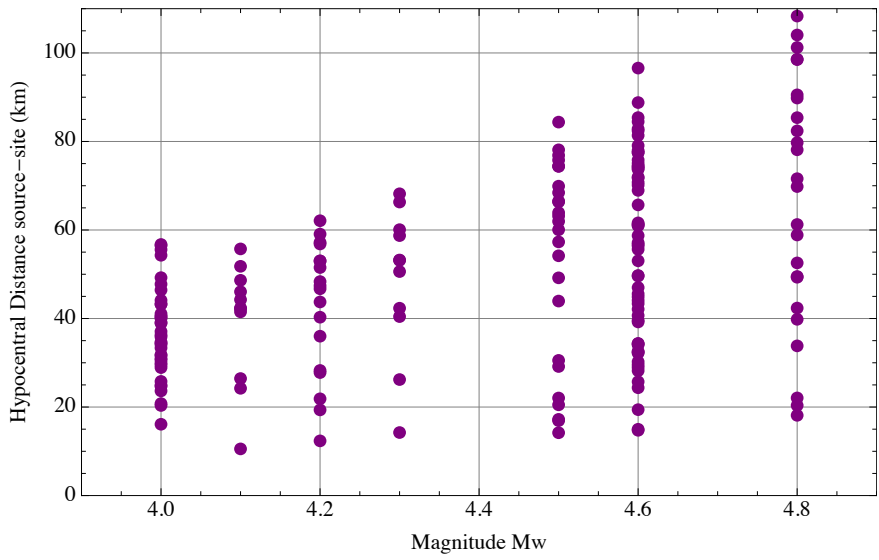

**Fig. 10**

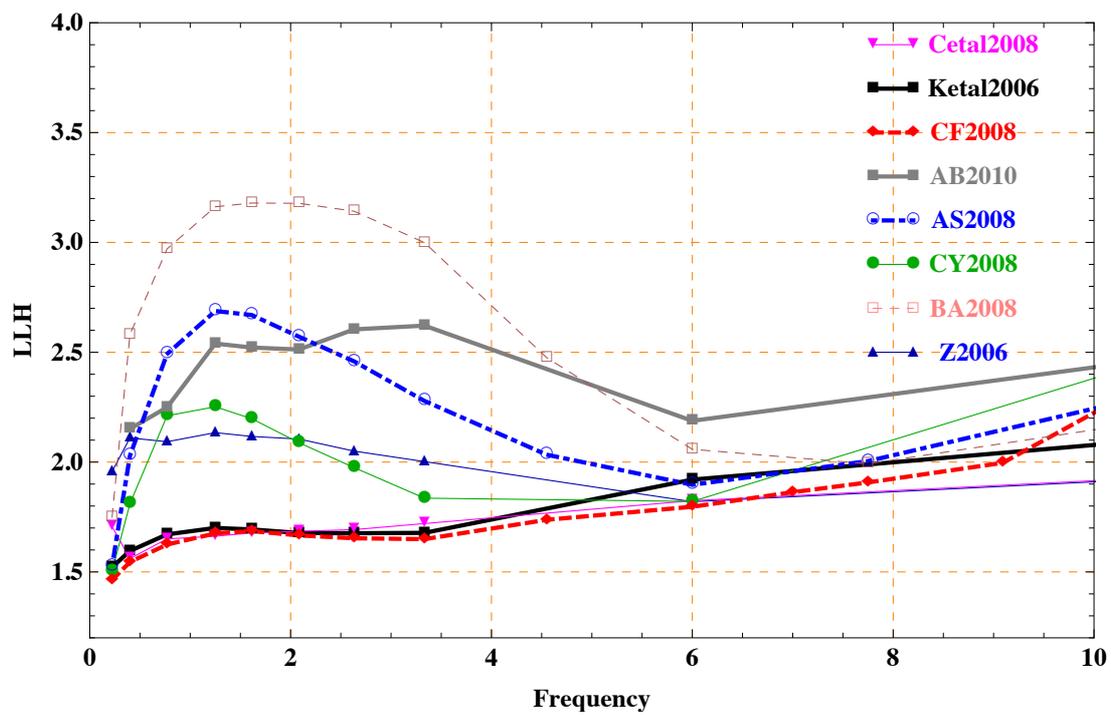

**Fig. 11**

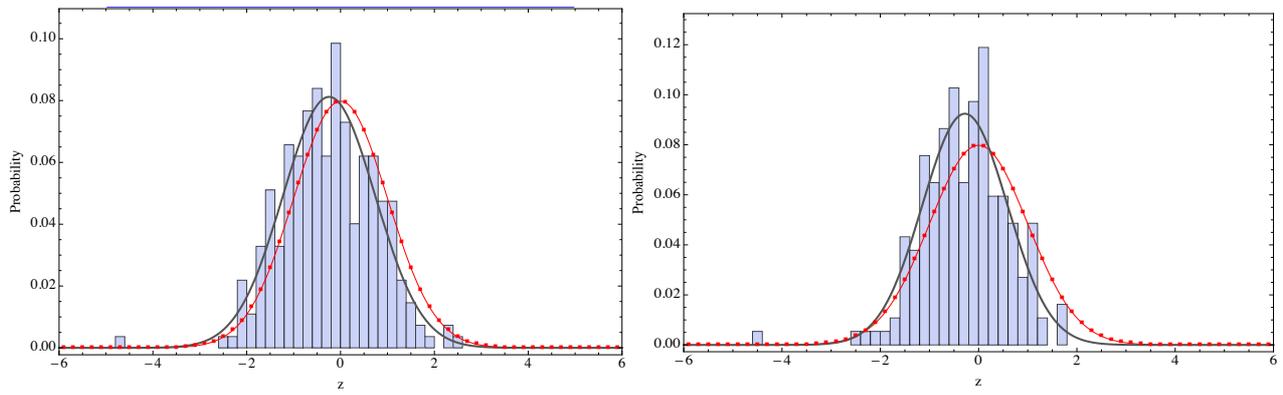

**Fig. 12**